\documentclass[letterpaper,nofootinbib]{revtex4}
\usepackage{amsmath}
\usepackage{amsthm}
\usepackage{bm}
\usepackage{graphicx}
\usepackage[flushleft]{caption}
\usepackage{subfig}
\usepackage{longtable}
\usepackage[dvips]{hyperref}
\begin{document}
\title{PAMELA Satellite Data as a Signal of Non-Thermal Wino LSP Dark Matter}
\author{Gordon Kane}
\email{gkane@umich.edu}
\author{Ran Lu}
\email{luran@umich.edu}
\author{Scott Watson}
\email{watsongs@umich.edu}
\affiliation{Michigan Center for
Theoretical Physics, University of Michigan, Ann Arbor, Michigan 48109, USA}

\begin{abstract}
Satellite and astrophysical data is accumulating that suggests
and constrains interpretations of the dark matter of the universe. We argue
there is a very well motivated theoretical framework (which existed before
data) consistent with the
interpretation that dark matter annihilation is being observed by the PAMELA
satellite detector. The dark matter is (mainly) the neutral W boson superpartner,
the wino. Using the program GALPROP extensively we study the annihilation
products and the backgrounds together. A wino mass approximately in the
$180-200 \,\mathrm{GeV}$ range gives a good description of the PAMELA data,
with antimatter and gammas from annihilating winos dominating the data below
this energy range but not contributing above it. We explain why PAMELA data
does not imply no antiproton signal was observed by PAMELA or earlier
experiments, and explain why the antiproton analysis was misunderstood by
earlier papers. Wino annihilation does not describe the Fermi
$e^+ + e^-$ data (except partially below $\sim  100 \,\mathrm{GeV}$). At higher energies
we expect astrophysical mechanisms to contribute, and we simply parameterize
them without a particular physical interpretation, and check that the
combination can describe all the data. We emphasize several predictions for satellite data
to test the wino interpretation, particularly the flattening or
turndown of the positron and antiproton spectra above 100 GeV. It
should be emphasised that most other interpretations require a large rise in the
positron and antiproton rates above $100 \,\mathrm{GeV}$. We focus on studying
this well-motivated and long predicted wino interpretation, rather than
comparisons with other interpretations. We emphasize that interpretations
also depend very strongly on assumptions about the cosmological history of the
universe, on assumptions about the broader underlying theory context, and on
propagation of antiprotons and positrons in the galaxy. The winos PAMELA is
observing arose from moduli decay or other non-thermal sources rather than a
universe that cooled in thermal equilibrium after the big bang. Then it is
appropriate to normalize the wino density to the local relic density, and no
``boost factors'' are needed to obtain the reported PAMELA rates.
\end{abstract}
\maketitle
\section{Introduction}
How does one learn what physics interpretation to give to tentative signals of
antimatter and gammas in the galaxy? Could it be due to annihilating dark
matter? Answering this is not straightforward -- it strongly depends on
assumptions that are not always made explicit, and the answer is also very
sensitive to assumptions about propagation in the galaxy, and to parameters
used to describe the propagation.
Perhaps surprisingly, assumptions about cosmological history are crucial. It
also depends on whether more than one mechanism is providing the signals.

As the recent PAMELA and then Fermi satellite data appeared, essentially
everyone who studied it assumed that the universe cooled in thermal
equilibrium after the big bang (dark matter particles $\chi $ annihilated into
Standard Model particles, which could annihilate back into dark matter
particles if they had enough energy, until the cooling led to freeze-out of
the dark matter at some relic density). Then the relic density is $\rho
\approx H/\langle \sigma v\rangle $, where the Hubble parameter $H$ is
evaluated at the freeze-out temperature (about $M_{\chi }/25)$. This
remarkable formula, with the relic density depending only on the cosmological
Hubble parameter and on the weak scale annihilation rate, has been called the
"wimp miracle". Getting the correct relic density implied that $\langle
\sigma \upsilon \rangle \approx 3\times 10^{-26}\, \mathrm{cm^{3}s^{-1}}$, so any
candidate with a larger annihilation cross section was excluded. In recent
years, however, it has become increasingly clear that comprehensive underlying
theories which explain more than one thing at a time generically have
additional sources of dark matter, such as decaying particles, and therefore
that assuming thermal equilibrium as the universe cools is oversimplified and
misleading. This was noticed in
\cite{Kamionkowski:1990ni,Chung:1998ua,Giudice:2000ex},
emphasized a decade ago by Moroi and Randall\cite{Moroi:1999zb}, and more recently documented
in detail in a model based on a string theory construction with M theory
compactified on a manifold with G$_{2}$ holonomy\cite{Acharya:2008bk}. See
also \cite{Kane:2008gb}.

Further, the standard assumption of most studies was that a single candidate
had to describe the data for electrons, positrons, antiprotons, and gammas at
all energies. It's not clear why one would assume that, since in the presence
of dark matter there will in general be contributions from the annihilating
dark matter, and also from several astrophysical sources such as interstellar
medium accelerated by supernova remnant shock waves, pulsars, and perhaps
more. In some theories the dark matter is metastable, its decay induced by
much higher dimension operators, and contributes \textit{both} via
annihilation and via decay (work in progress)

Dark matter composed of the lightest superpartner has been a very well
motivated candidate for all or part of the dark matter of the universe for nearly three
decades. The dark matter annihilation cross section is of order $\sigma \sim
\alpha _{2}/M_{\chi }^{2}$, where $\alpha _{2}$ is the weak coupling. In
the past decade the particular possibility where the dark matter is mainly
the partner of the W boson, the wino, has been very well motivated in any
theory where supersymmetry is broken by the anomaly mediation mechanism
\cite{Randall:1998uk}, or more generally where the anomaly mediation contribution to gaugino
masses is comparable with other sources of gaugino masses (as in the G$_{2}$
construction), and in other approaches such as U(1) mediation
\cite{Langacker:2007ac}. In a
universe where the relic density emerged in thermal equilbrium the wino mass
had to be of order $2$ TeV to get the right relic density. But in the
non-thermal universes where the dark matter mainly arose from decay of
additional particles, such as the moduli generically present in any string
theory, the correct temperature at which to evaluate the Hubble parameter
was the moduli decay or reheating temperature. This is quite different from the
freeze-out temperature \cite{Moroi:1999zb,Acharya:2008bk,Kane:2008gb}. Then the correct relic
density emerged for wino masses of order 200 GeV, for which $\langle \sigma
\upsilon \rangle \simeq 3\times 10^{-24}\,\mathrm{cm^{3}s^{-1}}$. Although the winos
arise continuously as the moduli decay, rather than the superpartners being
mainly present at the big bang, a "non-thermal wimp miracle" still occurs
when the scaling of the Hubble parameter and cross section with temperature
and mass are taken into account \cite{Moroi:1999zb,Acharya:2008bk,Kane:2008gb}.

Remarkably, this 200 GeV mass scale is just the one that is right for the
PAMELA data. In a universe where the relic density arises non-thermally,
as generically in string theories, a wino LSP with relic density normalized
to the observed local relic density ($0.3 \, \mathrm{GeV/cm^{3}}$) gives about the
amount of positrons and antiprotons (and their distributions) reported in
the PAMELA experiment! No "boost factor" is needed.

Positrons and antiprotons have long resident times in the galaxy, millions
of years. In order to compute the number of events as functions of energy
that PAMELA and Fermi should observe one needs to include all the effects of
propagation in the galaxy. There are two main programs to facilitate that,
GALPROP\cite{Strong:1998pw} and DarkSUSY\cite{Gondolo:2004sc,DarkSUSYWebPage}, each valuable for somewhat different calculations. \
Here we use GALPROP since we need to have one program that treats the signal
and background particles in a self-consistent manner as they are affected by the galactic
magnetic fields, lose energy by synchrotron radiation, inverse Compton
scattering, collisions, escape the galactic disk, etc. As we will explain,
this is crucial for understanding the antiprotons, where we argue that use
of parameterized backgrounds has led people incorrectly to assume that
PAMELA was not seeing an annihilation signal in antiprotons.

We find that in the PAMELA region the results can depend significantly on a
number of astrophysical parameters (see Table \ref{tab:param}), and that there are many
degeneracies and flat directions among the parameters. The GALPROP running
time is long, of order several hours, so we have not yet been able to do full
parameter scanning. Improved computing, and additional constraints from
satellite data that should be reported in the next few months, should improve
this situation significantly. It should be emphasized that the positrons and
antiprotons "injected" into the galaxy by pure wino annihilation have no
parameters apart from the mass scale, which can only vary at the 5-10\% level.
All the issues about describing the data arise from the propagation.

If annihilation of LSP dark matter is the origin of the excess positrons, it
obviously must give excess antiprotons since all MSSM states will include
quarks and antiquarks in their annihilation products, and the antiquarks
fragment into antiprotons. In particular, for the wino LSP the annihilation
of winos is to $W^{+} + W^{-}$, and the W-bosons have known branching ratios
to leptonic and quark final states, and the probability they will give
antiprotons was measured at LEP. The relevant processes are incorporated
into PYTHIA and we use them. There is no freedom. This has led to many
statements in literature that the apparent absence of an antiproton excess
excludes MSSM LSP models, and in particular excludes wino annihilation as the
explanation of the positron excess, and forces one to approaches that only
give leptons. It turns out that these conclusions are wrong, for three
interesting reasons. First, the antiproton spectrum from quark fragmentation
is significant down to quite soft antiprotons, and it gives a significant
number of antiprotons in the $1-10 \,\mathrm{GeV}$ region and even below. \
The positron spectrum from W's has many energetic positrons at higher
energies, and is peaked at higher energies than the antiproton spectrum. \
Second, the antiprotons do not lose much energy as they propagate compared to
the positrons, so the GeV antiprotons are detected by PAMELA, while the
positrons lose energy readily and the soft ones do not make it to the
detector. This can be seen from the figures below, where the signal from
antiprotons is above the background down to the lowest energies, while the
positron signal is at the background level below about $5 \,\mathrm{GeV}$, and
essentially gone below $10 \,\mathrm{GeV}$. Thus the positron spectra for signal and background have different shapes,
while the antiproton spectra have essentially the same shape and mainly differ
in normalization.

The third issue concerns how the background is defined. The "background" \
can only be defined if one either has data in a region where there is known to
be no signal, or if one has a theory of the background. There are two
points. Some time ago reference \cite{Donato:2003xg} showed that solving the propagation
equations allowed a rather large variation in the antiproton background
normalization, about a factor of five for parameters that were consistent with
other constraints such as the Boron/Carbon ratio. This has also been
emphasized by \cite{Donato:2003xg,Bergstrom:2006tk}. Low energy data for
antiprotons has existed for over a
decade\cite{Orito:1999re,Maeno:2000qx,Asaoka:2001fv}. People proceeded by fitting
the antiproton data and defining the fit as the background. Then they compared
the PAMELA data with such backgrounds, and of course concluded there was no
signal, because the signal had already been included in the background! If a
dark matter annihilation contribution was included, it was double counted! \
As we show below, for entirely reasonable GALPROP parameters one can
self-consistently compute the antiproton background, and with a wino
annihilation signal it gives a good description of the data. Consequently
the early experiments such as BESS and HEAT \textit{did} detect dark matter
via antiprotons.

There is another effect that has been neglected so far in most interpretations
of the PAMELA and Fermi data. Dark matter annihilation is proportional to
the square of the relic density. Because galaxies are built from smaller
galaxies, and also because of normal random density fluctuations, the relic
density throughout the relevant parts of the galaxy for a given observable
will not be a flat $0.3 \,\mathrm{GeV/cm^{3}}$, but will vary. Since
$\left\langle \rho ^{2}\right\rangle -\left\langle \rho \right\rangle ^{2}\neq
0,$ density fluctuation effects must occur. Initial studies have been done
by several authors\cite{Lavalle:2006vb,Lavalle:1900wn}, who have established that the effects differ for
positrons and for antiprotons, and are energy dependent, because the energy
loss mechanisms are significantly different for positrons and antiprotons. \
While it is not clear yet how to calculate accurately the sizes and energy
dependences of these effects, it is likely that assuming no effect is a less
good approximation than initial approximate calculations of the effects. We
include small effects we estimate semi-analytically and show results with and
without them. Ultimately it will be very important to learn how to calculate
these effects well and include them.

The positron data below $10-15 \,\mathrm{GeV}$ is not consistent among experiments, and
is not well described by models. This is assumed to be due to
charge-dependent solar modulation effects and is being actively studied by
experts in that area, and by the experimenters
\cite{1996ApJ...464..507C,RomePAMELA}. We do
not attempt to put in detailed corrections for this, but we do include the
(non-charge-dependent) effects in the simulation. Results in
general also depend on the profile of the dark matter in the galaxy, most
importantly for gammas from the galactic center. We do not study this
dependence much here since it does not much affect the positron and antiproton
results, and we are computer-limited. We remark on it in context below.

The PAMELA experiment has reported deviations from expected astrophysics for
positrons, and as we explained above, for antiprotons, below about $100
\,\mathrm{GeV}$. In this paper our goal is to demonstrate that a wino LSP is a strong candidate
for explaining these deviations. We assume that
possible astrophysical sources can be added to give a complete description of
the data including Fermi (which light wino annihilation obviously cannot
explain). We only parameterize the higher energy
astrophysical part, and assume it can be accommodated by some combination of
acceleration of the interstellar medium electrons by supernova remnant shock
waves, pulsars, etc with a net $e^+/e^-$ ratio of 1/6. Once we do that we make a number of predictions for the positron
ratio and antiprotons at energies above those already reported, for diffuse
gammas and gammas from the galactic center, and for gamma fluxes from dwarf
galaxies in our galactic halo. We check constraints from synchrotron radiation and
"WMAP" haze.
Our prediction for gammas assumes the source of higher energy $e^{+}+e^{-}$
does not also produce a significant flux of high energy gammas. So corrections may be needed here,
but the gammas from wino annihilation will be an irreducible source.

We also do not criticize any other attempt to describe the data -- in nearly
all cases forthcoming PAMELA and Fermi data will favor one or another approach. Many are based
on interesting ideas and models. While some predictions are very sensitive to
propagation effects, others such as whether the positron ratio rises or falls
above $100 \,\mathrm{GeV}$ is not very sensitive to propagation and different
models have very different predictions.  We predict a fall or flattening
(depending somewhat on the high energy astrophysical component), while most
models predict a strong rise. While propagation uncertainties do not modify
our qualitative predictions, it is clear that the presence of the rather hard
contribution to the $e^{+}+e^{-}$ flux introduces uncertainty in our
predictions, since it could affect the positron ratio at lower energies, and
it might or might not contribute to antiproton and gamma fluxes. As
described below, we simply parameterize it and we assume it only contributes
to the electrons and positrons, and that it contains mostly electrons
($e^{+}/e^{-}=1/6)$. This limits the quality of our predictions.

In the following we first describe our use of GALPROP in some detail,
including some effects or constraints we incorporate. Then we show the data
and a description of the data based on annihilation of $180 \,\mathrm{GeV}$
dark matter winos plus cosmic ray backgrounds, with signal and backgrounds
computed in a consistent manner. What we show is not a fit to data but
merely educated guesses since the computing time for a full parameter scan is
still prohibitive for us. As shown in Table \ref{tab:param} below, we vary
eight
GALPROP parameters and some others, and all of them affect the
interpretation. We have established that comparable descriptions of the data
can be obtained for different GALPROP parameters from those we show since
there are degeneracies. We emphasize that our goal is to show that the wino
LSP in the mass range of order $200 \,\mathrm{GeV}$ is a good candidate for
the dark matter of the universe, including theoretical and experimental
information as well as it is possible today. Even though much is not known
about the propagation of the electrons, positrons, antiprotons and
gammas, quite a lot \textit{is} known. Then we describe the contribution we
arbitrarily assume for the higher energy electrons and positrons. We will
report on studies of the GALPROP parameter degeneracies and the wino mass
later, assuming our predictions for the positron and antiproton higher energy
data are correct.

After that we turn to presenting the data and the descriptions. We do that
for one mass and one set of propagation parameters, and one dark matter
profile (NFW). Descriptions and
predictions for higher energies are shown for the PAMELA positron excess, the
PAMELA antiproton excess, the $e^{+}+e^{-}$ sum, the diffuse gammas, the
gammas from the galactic center, dwarf galaxies, and checks such as the
Boron/Carbon ratio. Finally we present conclusions and a list of tests of
the wino LSP model. Definitive tests of this approach will also occur at the
LHC; we will present those elsewhere. When future data is reported we will
post updated graphs at \url{http://wino.physics.lsa.umich.edu}.

\bigskip

\section{GALPROP Parameters}

We use GALPROP v50.1p\cite{Strong:1998pw} to simulate the propagation of both cosmic rays
and dark matter annihilation products in the galaxy in a consistent way. The
rates and spectrum for both will change if the propagation parameters change.
The propagation process is described by the propagation equation for the
particle density $\psi$:

\begin{align}
\frac{\partial \psi }{\partial t}=q\left( \vec{r},p\right) & +\vec{\nabla}%
\cdot \left( D_{xx}\vec{\nabla}\psi -\vec{V}\psi \right) +\frac{\partial }{%
\partial p}p^{2}D_{pp}\frac{\partial }{\partial p}\frac{1}{p^{2}}\psi  
\notag \\
& -\frac{\partial }{\partial p}\left[ \dot{p}\psi -\frac{p}{3}\left( \vec{%
\nabla}\cdot \vec{V}\right) \psi \right] -\frac{1}{\tau _{f}}\psi -\frac{1}{%
\tau _{r}}\psi   \label{eqn:propagation}
\end{align}%
where $D_{xx}$ is the diffusion constant which is determined by: 
\begin{equation}
    D_{xx}=\beta D_{0xx}\left( \frac{\mathcal{R}}{\mathcal{R}_{0}}\right) ^{\delta }
\label{eqn:diffcoeff}
\end{equation}%
where $\beta $ is the velocity of the particle, $\mathcal{R}$ is the particle
rigidity, and $\mathcal{R}_{0}$ is the reference rigidity which is taken to be
$4\, \mathrm{GV}$ in all the simulations. In order to consider the propagation
of the dark matter signals in the same framework, the official GALPROP code is
modified to accept the dark matter injection spectrum calculated using PYTHIA
via
DarkSUSY 5.0.4\cite{Gondolo:2004sc}. The parameters used in this paper are based on the
conventional model with constant Xco-factor provide in GALPROP source code
(galdef\_50p\_599278). In general we vary the parameters $D_{xx}$, $\delta $ ,
the half height of the diffusion zone $z_{h}$, the primary electron injection
index $\gamma _{0}$, the normalization of the primary electron flux
$N_{e^{-}}$, the scaling factor for inverse Compton scattering, the convection
velocity $V_{c}$, and the Alfv\'{e}n velocity $V_{a}$. We survey ranges of
these parameters (but do not fit data or scan parameters because of computing
limitations) in order to learn if a combination of conventional cosmic ray
physics plus dark matter annihilation can give a reasonable description of the
PAMELA and Fermi data within certain constraints such as the B/C ratio
described further below.

We find a set of parameters that give a good description of the data, and
those are used in the figures below except when stated otherwise.The half
height of the diffusion zone $z_h = 2 \,\mathrm{kpc}$. The diffusion
coefficient is $D_{0xx} = 2.5\times 10 ^{28} \,\mathrm{cm^2 s^{-1}}$, $ \delta
= 0.5$. We also assume a softer primary electron injection spectrum by setting
the injection index $\gamma_0$ for primary electrons between $4 \,
\mathrm{GeV}$ to $10^6 \,\mathrm{GeV}$ to $2.6$. Also the primary electron
flux is normalized to $N_{e^-} = 2.88\times10^{-10} \,\mathrm{cm^{-2} sr^{-1}
s^{-1} MeV^{-1}}$ at $34.5 \,\mathrm{GeV}$. The scaling factors for inverse
Compton (ISRF factor) are adjusted to $0.5$, which is approximately equivalent
to setting $\tau = 2\times10^{16} \, \mathrm{s}$ in the energy loss rate
formula for electron $b\left( \varepsilon \right)_{e^{\pm}} =
\frac{1}{\tau}\varepsilon^2$. The convection velocity is $V_c = 5 \,\mathrm{
km\, s^{-1} kpc^{-1}}$, and the Alfv\'en speed which determines the reacceleration
process is $V_a = 31 \,\mathrm{km\, s^{-1}}$. The summary of the parameters we
are using can be found in Table \ref{tab:param}. Other parameters not
mentioned here are the same as galdef\_50p\_599278.

As described in the introduction, the dark matter we focus on is the
theoretically
well-motivated case of a pure wino LSP, which annihilates to $W$'s: $
\widetilde{W} + \widetilde{W} \rightarrow W^+ + W^-$. The dark matter
injection parameters are then completely determined by the wino mass and $W^{\pm}$ decays.

\begin{table}
\centering
\begin{tabular}{|c|c|}
\hline
\multicolumn{2}{|c|}{GALPROP Parameters} \\ \hline
$D_{xx0} \left( \mathrm{cm^2 s^{-1}} \right)$ & $2.5\times10^{28}$ \\ \hline
$\delta$ & $0.5$ \\ \hline
$\mathcal{R}_0 \left( \mathrm{GV }\right)$ & $4$ \\ \hline
$z_h \left(\mathrm{kpc}\right)$ & $2$ \\ \hline
$\gamma_0$ & $2.6$ \\ \hline
$N_{e^-} \left( \mathrm{cm^{-2} s^{-1} sr^{-1} MeV^{-1}} \right)$ & $%
2.88\times10^{-10}$ \\ \hline
$V_c \left( \mathrm{km s^{-1} kpc^{-1}} \right)$ & $5$ \\ \hline
$V_a \left( \mathrm{km s^{-1}} \right)$ & $31$ \\ \hline
ISRF factors (optical, FIR, CMB) & $0.5, 0.5, 0.5$ \\ \hline
\multicolumn{2}{|c|}{Solar Modulation Parameters} \\ \hline
$\phi \left( \mathrm{MV }\right)$ & $500$ \\ \hline
$p_c \left( \mathrm{GeV }\right)$ & $1$ \\ \hline
\multicolumn{2}{|c|}{Astrophysical Flux Parameters} \\ \hline
$a$ & $1.0$ \\ \hline
$b$ & $1.8$ \\ \hline
$z_0 \left( \mathrm{kpc }\right)$ & $0.2$ \\ \hline
$\gamma$ & $1.5$ \\ \hline
$M\left( \mathrm{GeV }\right)$ & $950$ \\ \hline
\multicolumn{2}{|c|}{Density Fluctuation Factor Parameters} \\ \hline
$B_c$ & $2.5$ \\ \hline
$f$ & $0.5$ \\ \hline
\end{tabular}
\caption{The parameters used for simulation. The physical meaning of these
parameters is described in the text.}
\label{tab:param}
\end{table}

\section{Solar Modulation}

The effect of solar modulation is estimated by the Force-Field approximation
\cite{1967ApJ...149L.115G}, The flux observed at Earth's orbit $J_{E}\left( \varepsilon \right)
$ is related to the flux in the interstellar flux by the following relation:
\begin{equation}
J_{E}\left( \varepsilon \right) =\frac{\varepsilon ^{2}-m^{2}}{\varepsilon
_{\infty }^{2}-m^{2}}J_{\infty }\left( \varepsilon _{\infty }\right) 
\label{eqn:modulation}
\end{equation}%
where $\varepsilon _{\infty }$ is the energy of the corresponding interstellar
flux, which is determined by:
\begin{equation}
\varepsilon _{\infty }=\left\{ 
\begin{array}{ll}
p\log \left( \frac{p_{c}+\varepsilon _{c}}{p+\varepsilon }\right)
+\varepsilon +\Phi  & \varepsilon <\varepsilon _{c} \\ 
\varepsilon +\Phi  & \varepsilon \geq \varepsilon _{c}%
\end{array}%
\right.   \label{eqn:energy}
\end{equation}%
$\Phi $ is the modulation energy shift which can be calculated from the
modulation potential $\Phi =\left\vert Z\right\vert e\phi $, In our simulation
the modulation potential $\phi $ is $500\,\mathrm{MV}$, the reference momentum
$p_{c}$ is $1\,\mathrm{GeV}$. Only the solar modulation effect for
electron/positron and antiproton/proton is considered in this work. These
effects do not include charge dependent solar modulation, which is under study
\cite{1996ApJ...464..507C,RomePAMELA}

\begin{figure}[htpb]
\centering
\includegraphics[width=15cm]{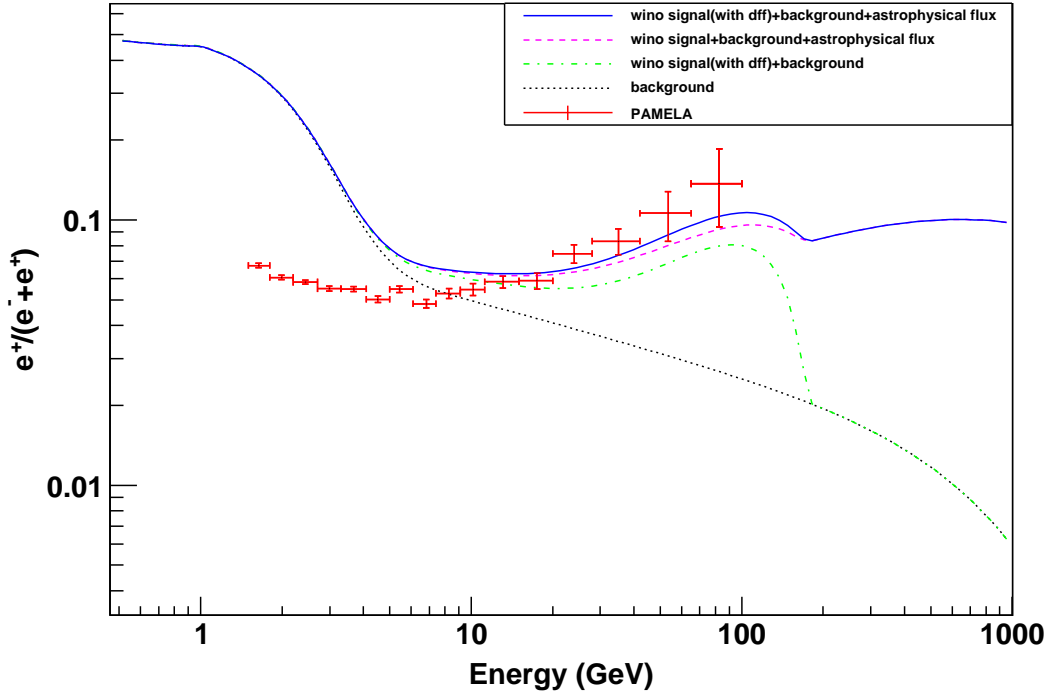}
\caption{The positron flux ratio, generated with the parameters described in
the text and Table \ref{tab:param} with a $M_{\widetilde{W}} = 180 \,
\mathrm{GeV}$ wino. The solid line is the ratio of the total positron flux,
which includes the positrons from the wino annihilation, the density
fluctuation factor, the astrophysical flux and the conventional astrophysics
background to positrons plus electrons The dash line has the same components
but without the density fluctuation factor. The dash-dot line contains just
the wino annihilation and the conventional astrophysics background, and the
dot line is the ratio of the secondary positrons only.The data are from
\cite{Adriani:2008zr}, Our analysis assumes the reported normalization of the Fermi and
PAMELA data. If those change it will affect the higher energy extrapolation
here. Note that the predicted positron fraction does not continue to rise. At
the PAMELA meeting in Rome\cite{RomePAMELA} data was reported with the four
higher energy points about 10\% lower than shown here, but we do not show that data since it
has not yet been published.}
\label{fig:ep_ratio}
\end{figure}

\begin{figure}[htpb]
\centering
\includegraphics[width=15cm]{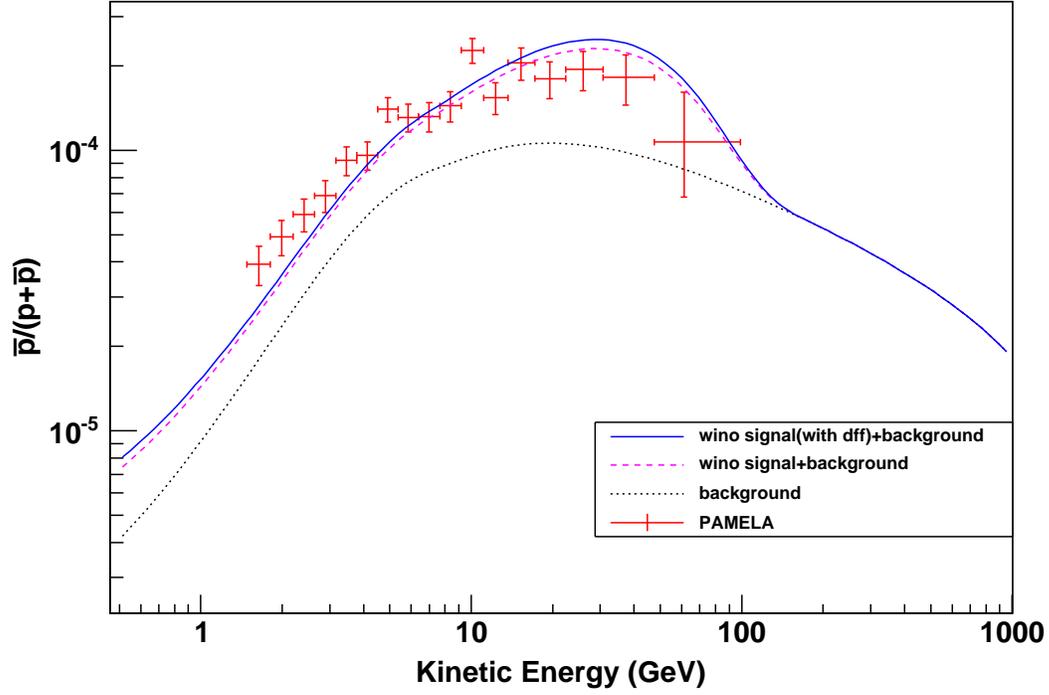} 
\caption{The antiproton flux ratio. The solid line is the ratio of the total
antiproton flux, which include the antiproton from wino annihilation, and
conventional astrophysics background, the dash line has the same components
but without the density fluctuation factor, the dot line is astrophysics
background only. The data are from PAMELA \cite{Adriani:2008zq}. At the PAMELA
meeting in Rome\cite{RomePAMELA}, data was reported with the last bin
increased by 70\%, and a bin up to $185 \,\mathrm{GeV}$ with three events, but we do
not show the data since it is not published. Note the signal is larger than
the background down to very low energies.}
\label{fig:pb_ratio}
\end{figure}

\begin{figure}[htpb]
\centering
\subfloat{\includegraphics[width=15cm]{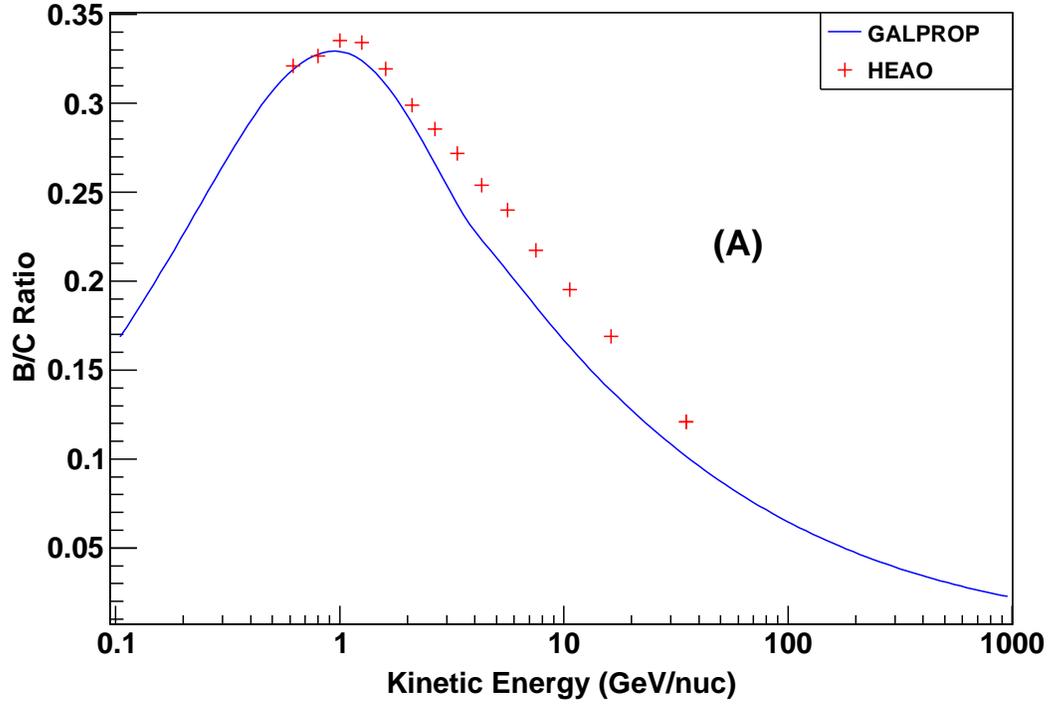}}
\caption{The Boron to Carbon ratio with our standard parameters, solar
modulation effect is not included, The data are from
\cite{1990A&A...233...96E}.}
\label{fig:B_C}
\end{figure}

\begin{figure}[htpb]
\ContinuedFloat
\centering
\subfloat{\includegraphics[width=15cm]{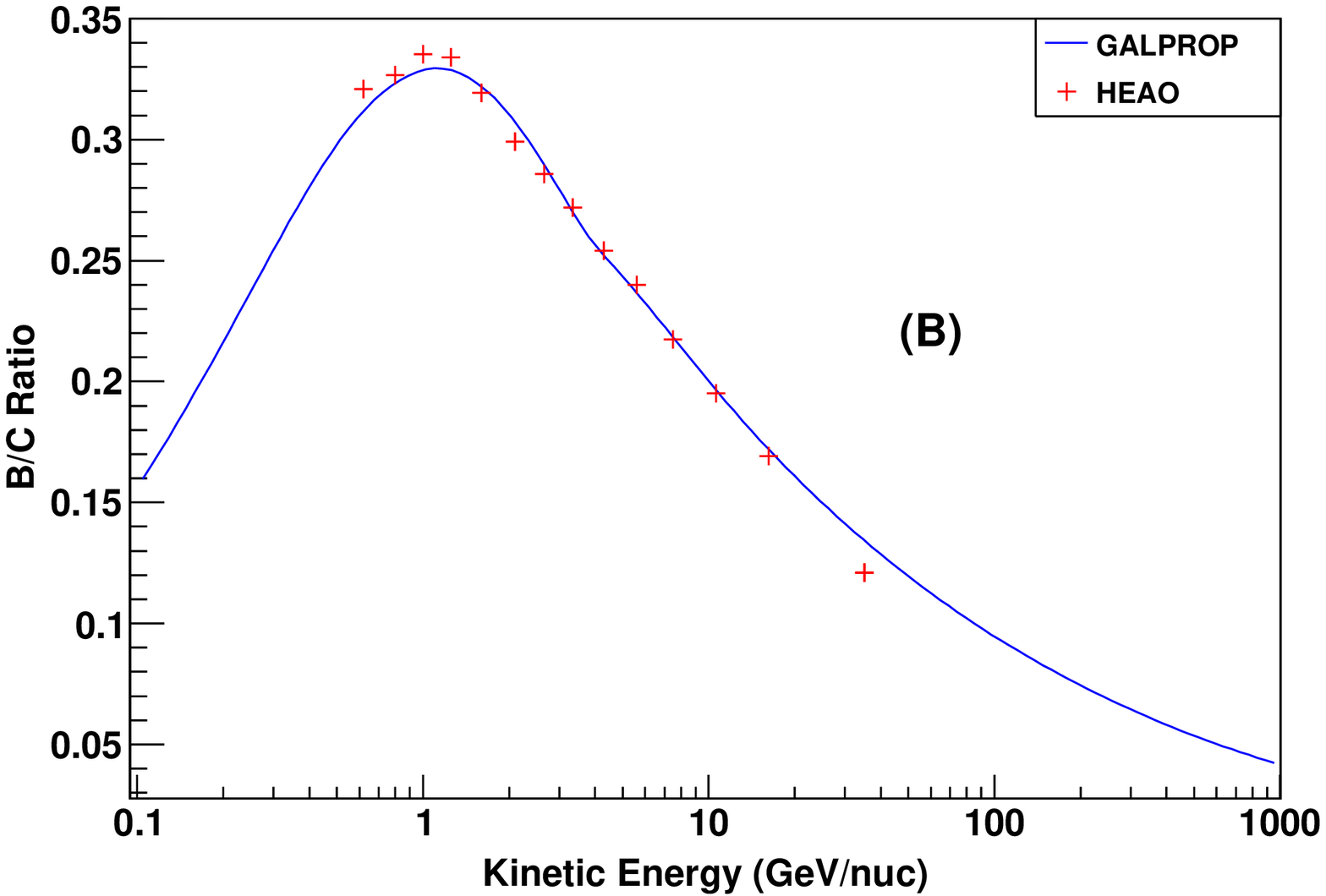}}
\caption{The Boron to Carbon ratio with one parameter different ($\delta$
changes from 0.5 to 0.4). This
illustrates that the Boron to Carbon ratio is very sensitive the diffusion
parameters. The data are from \cite{1990A&A...233...96E}.}
\label{fig:B_C1}
\end{figure}
\clearpage
\section{Astrophysical Flux}

It is obvious that a $180\,\mathrm{GeV}$ wino alone cannot explain Fermi date
and PAMELA data at the same time. There must be some extra flux responsible
for the high energy signals. In order to estimate the high energy
($>200\,\mathrm{GeV}$) electrons/positrons flux, we consider a simple model
for extra flux which is suggested by interstellar
medium electrons accelerated by supernova remnants and shock waves, or by
pulsar spectra models. The basic
setup we use is similar to Zhang and Cheng\cite{2001A&A...368.1063Z}. For a
recent review of pulsar models, see Profumo\cite{Profumo:2008ms}. The spatial distribution
of the sources is:
\begin{equation}
\rho \left( r\right) = N \left( \frac{r}{r_{\odot }}\right) ^{a}e^{-\frac{%
b\left( r-r_{\odot }\right) }{r_{\odot }}}e^{-\frac{z}{z_{0}}}
\label{eqn:pular}
\end{equation}%
where $N$ is the overall normalization constant, $z_{0}=0.2\mathrm{kpc}$,
$r_{\odot }=8.5\mathrm{kpc}$, $a=1.0$ and $b=1.8$.

The energy dependence of the injection spectrum is: 
\begin{equation}
\frac{dN_{e^{\pm }}}{dE}=N' E^{-\gamma }e^{-\frac{E}{M}}  \label{eqn:pular2}
\end{equation}%
where $N'$ is another normalization constant, which can be absorbed into $N$,
$\gamma =1.5$ and $M=950 \,\mathrm{GeV}$. In order to fit PAMELA and Fermi
data, we made an ad hoc assumption that the ratio of positron and electron in
the unknown extra flux is $1:6$. The high energy positrons/electrons are then
propagated by GALPROP, and the resulting flux is normalized to fit the Fermi
data by requiring the extra electron flux at $275.5 \, \rm GeV$ to be $3.0
\times 10^{-13} \,\mathrm{cm^{-2} sr^{-1} s^{-1} MeV^{-1}}$

\begin{figure}[htpb]
\centering
\includegraphics[width=15cm]{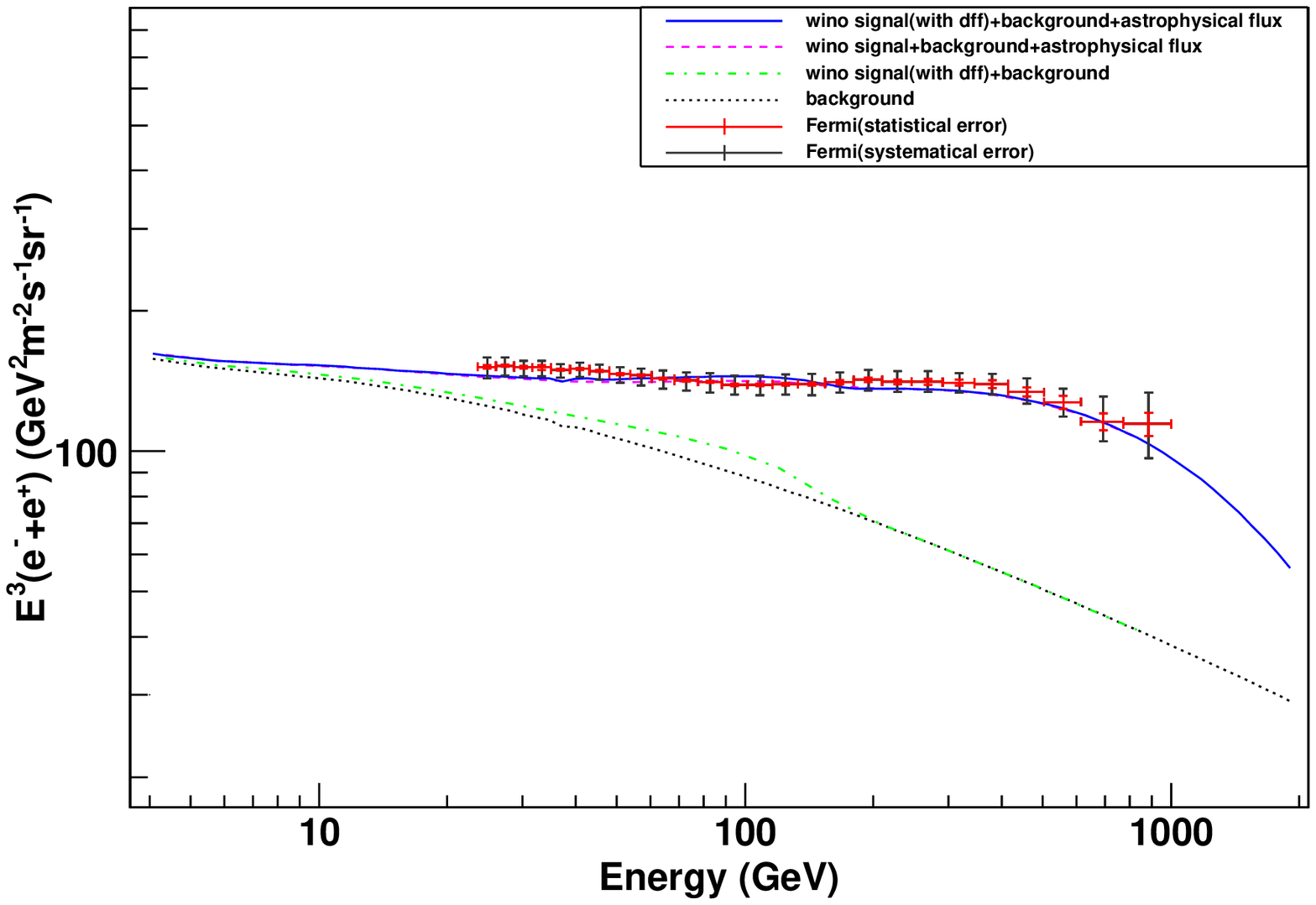} 
\caption{The absolute flux of $e^+ + e^-$, The solid line is the sum of
electron and positron from the wino annihilation, the density fluctuation
factor, our assumed extra flux, and conventional astrophysics background, the dash line
has the same components but without the density fluctuation factor. The
dash-dot line contains wino annihilation and astrophysics background, and the
dot line is the conventional astrophysics background only. See comments in
Figure \ref{fig:ep_ratio}. The data are from \cite{Abdo:2009zk}.}
\label{fig:e_flux}
\end{figure}

\begin{figure}[htpb]
\centering
\includegraphics[width=15cm]{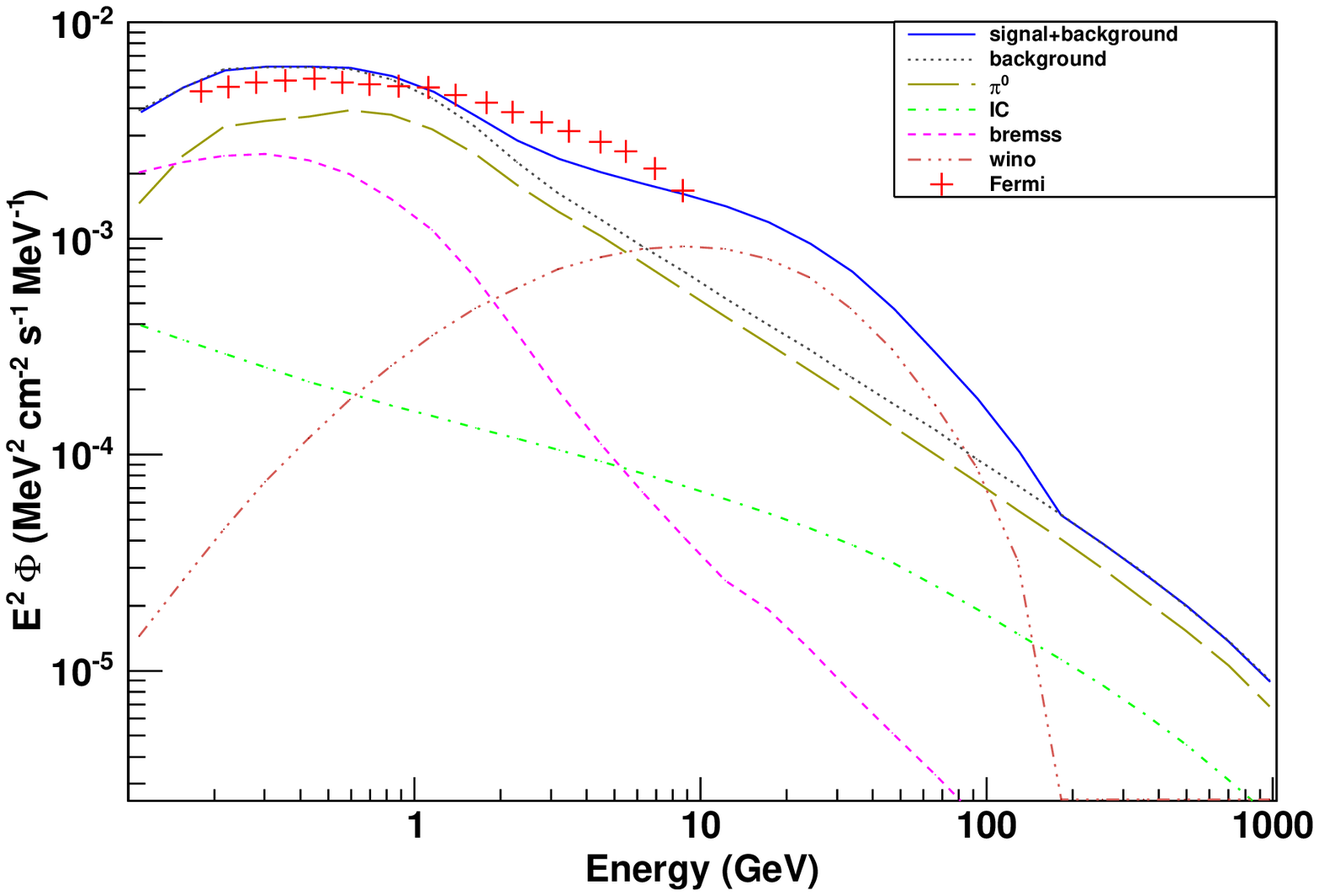} 
\caption{Different components of the diffusive gamma ray emission averaged
over $10^{\circ} \le \left|b\right| \le 20^{\circ}$ region, The gamma ray
emission is integrated over a spherical halo with radius $r = 20 \,
\mathrm{kpc}$. All parameters as in Table \ref{tab:param}, and wino
mass $M_{\widetilde{W}} = 180 \, \mathrm{GeV}$. The solid line is the total
flux, the dot line is the total background, the long dash line is the flux
from $\pi^0$ decay, the dash-dot line is the
flux from inverse Compton, the dash line is the flux from  bremsstrahlung, and
the dash-dot-dot line is the flux from wino annihilation. All parameters as in
Table \ref{tab:param}, and wino mass $M_{\widetilde{W}} = 180 \,
\mathrm{GeV}$. The data are from \cite{Morselli:LHC}.}
\label{fig:gamma_fermi_components}
\end{figure}

\begin{figure}[htpb]
\centering
\includegraphics[width=15cm]{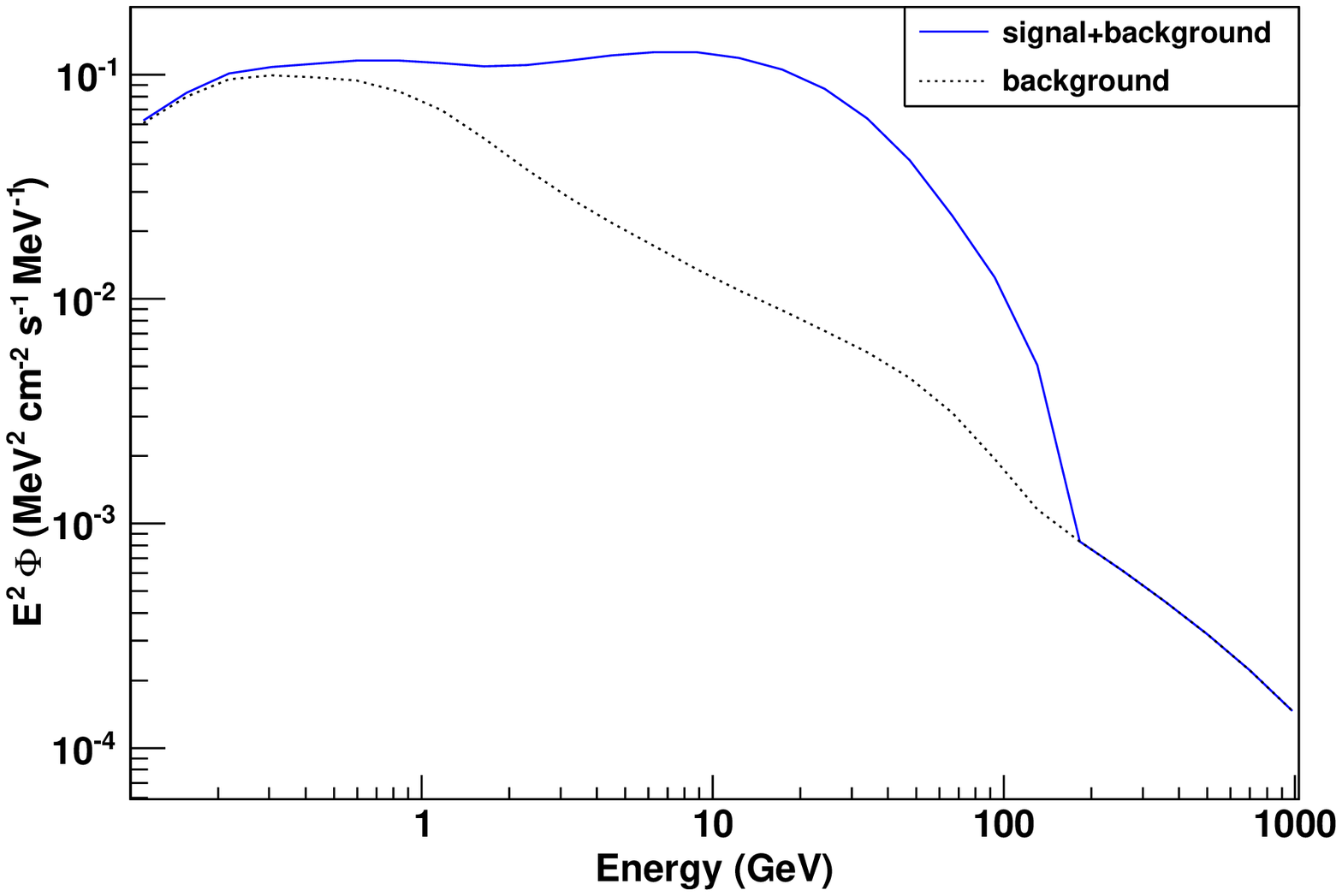} 
\caption{Gamma ray emission from the galactic center, the flux is averaged
over $\left|l\right| \le 0.5^\circ,\; \left|b\right| \le 0.5^\circ$. All
parameters as in Table \ref{tab:param}, and wino mass $M_{\widetilde{%
W}} = 180 \, \mathrm{GeV}$.}
\label{fig:gamma_center}
\end{figure}

\section{Density Fluctuation Factor}

The results from N-body simulations \cite{Diemand:2005vz}, and an understanding of how galaxies
formed, indicate that it is inevitable for the dark matter halo of our
galaxy to have substructures. The existence of these substructures would
change the predictions of the cosmic ray fluxes, particularly for dark matter
annihilation. Even without substructures, it is clear that the density of dark
matter will not be absolutely flat, but will fluctuate around an average
value. Both of these effects require the flux from dark matter annihilation,
which is sensitive to the square of the dark matter density, to show density
fluctuation effects. As emphasized by Lavalle and
collaborators\cite{Lavalle:2006vb,Lavalle:1900wn}, the
effects of these substructures are different for positrons and antiproton, and
must also be energy dependent. The details of these effects depend on the
spatial and mass distribution of these substructures. Here\footnote{The
analysis of this section was carried out in collaboration with Cheng Peng.} we use a very
simple model to estimate the effects: Assuming all the substructures share the
same mass and density, and the spatial number density of the substructures is
proportional to the density profile of the smooth distribution of dark matter
halo.

With these assumptions, the density fluctuation factor can be calculated with:
\begin{equation}
D\left( E\right) =\left( 1-f\right) ^{2}+fB_{c}\frac{\mathcal{I}_{1}}{%
\mathcal{I}_{2}}  \label{eqn:beff}
\end{equation}%
where $f$ is the mass fraction of the substructures in the dark matter halo,
$B_{c}$ is the intrinsic density increase of the substructures or
fluctuations, and $\mathcal{I}_{n}$ is determined by 
\begin{equation}
\mathcal{I}_{n}=\int_{\mathrm{DM\;halo}}G\left( x,E\right) \left\{ \frac{%
\rho _{s}\left( x\right) }{\rho _{0}}\right\} ^{n}d^{3}x
\label{eqn:integral}
\end{equation}%
where $G\left( x,E\right) $ is the Green's function of the propagating
particle.

We assume $f=0.5$ and $B_{c}=2.5$. Only the density fluctuations of positron
and antiproton are considered. The electron/positron Green's function from
Baltz and Edsj\"{o}\cite{Baltz:1998xv} and the fast formulae for antiproton Green's
function from Maurin, Taillet and Combet\cite{Maurin:2006hy} are used to evaluate the
integral. With these assumptions we find the effects are small. We included
them for completeness. Theory curves including them are labeled ``dff'' (for
``density fluctuation factors'') in the figures.

\section{Dwarf Galaxies}
Dwarf spheroidal satellite galaxies are unique targets for indirect
detection of dark matter\cite{Evans:2003sc}. Most of them are believed to be
dark matter dominated objects and some of them are located at high Galactic
latitudes, which reduces the diffusive gamma background. But the dark matter
density profiles of these dwarf galaxies are hard to determine, which gives
large uncertainties in the predictions\cite%
{Strigari:2006rd,Strigari:2007at,Essig:2009jx}.

We calculate the gamma ray flux from wino annihilation in these dwarf
galaxies following Essig, Sehgal and Strigari\cite{Essig:2009jx}. The
formula for the gamma ray flux from annihilating dark matter in a dark
matter halo is 
\begin{equation}
\frac{dN_{\gamma }}{dAdt}=\frac{1}{8\pi }\mathcal{L}_{\mathrm{ann}}\frac{%
\left\langle \sigma v\right\rangle }{M_{\widetilde{W}}^{2}}\int_{E_{\mathrm{%
th}}}^{E_{\max }}\frac{dN_{\gamma }}{dE_{\gamma }}dE_{\gamma }
\label{eqn:dwarf}
\end{equation}%
where in our model the annihilation cross-section $\left\langle \sigma
v\right\rangle =2.50\times 10^{-24}\,\mathrm{cm^{3}s^{-1}}$, and the wino
mass $M_{\widetilde{W}}$ is set at $177.5\,\mathrm{GeV}$ (it is not exactly
180 GeV because the soft SUSY breaking terms changed the spectrum a little
bit). We take the threshold energy $E_{\mathrm{th}}=100\,\mathrm{MeV}$, the
integration $\int_{E_{\mathrm{th}}}^{E_{\max }}\frac{dN_{\gamma }}{%
dE_{\gamma }}dE_{\gamma }=27.14$.  $\mathcal{L}_{\mathrm{ann}%
}=\int_{0}^{\Delta \Omega }\left\{ \int_{LOS}\rho ^{2}(r)ds\right\} d\Omega ,
$ which only depends on the properties of the dark matter halo and the solid
angle over which it is observed, can be found in Table 1 of \cite%
{Essig:2009jx}. The results of our estimate of the gamma ray fluxes from
several dwarf galaxies is presented in Table \ref{tab:dwarf}. Although
limits are reported for some dwarf galaxies \cite{Winer:susy09} we cannot
directly compare our predictions with those limits since they are somewhat
dependent on the analysis.  We estimated the corresponding constraints for
the wino LSP by scaling the cross section upper bound for the $b\bar{b}$
final state provided by \cite{Winer:susy09} to the corresponding wino LSP
cross section upper bound which gives the same amount of gamma ray flux.
Different dSph give upper bounds on the cross section ranging from $%
1.6\times 10^{-24}\,\mathrm{cm^{3}s^{-1}}$ to $3.6\times 10^{-23}\,\mathrm{%
cm^{3}s^{-1}}$, with Willman1 giving the most stringent constraint. As shown
in the Table, after the uncertainty of the halo density profile is taken
into account, the wino LSP predictions are allowed by the limits. \ At the
same time, the predictions are large enough to anticipate seeing a signal
soon.

\begin{table}[tbp]
\centering
\begin{tabular}{|c|c|c|}
\hline
Dwarf & $\mathcal{L}_{\mathrm{ann}}$ & Flux $\left(E_{\mathrm{th}} = 100 \,%
\mathrm{MeV} \right)$ \\ 
Galaxies & $\log_{10} \left[\mathrm{GeV^2 c^{- 4} c m^{- 5}}\right]$ & $%
10^{-9} \mathrm{cm^{- 2} s^{- 1}}$ \\ \hline
Sagittarius & $19.35 \pm 1.66$ & $1.9\; \left(0.042, 88\right) $ \\ \hline
Draco & $18.63 \pm 0.60$ & $0.36\; \left(0.092, 1.5\right)$ \\ \hline
Ursa Minor & $18.79 \pm 1.26$ & $0.53\; \left(0.029, 9.6\right)$ \\ \hline
Willman 1 & $19.55 \pm 0.98$ & $3.0\; \left(0.32, 29\right) $ \\ \hline
Segue 1 & $20.17 \pm 1.44$ & $13\; \left(0.46, 350\right)$ \\ \hline
\end{tabular}
\caption{The wino LSP prediction of gamma ray flux from some dwarf galaxies,
based on the calculation of $\mathcal{L}_{\mathrm{ann}}$ in \cite%
{Essig:2009jx}. The three numbers in the flux column are the central value
and the maximum and minimum value corresponding to the $\mathcal{L}_{\mathrm{%
ann}}$ result. The results are consistent with the recently reported
preliminary flux upper limits for 3-month Fermi LAT data \cite%
{Winer:susy09}, and suggest signals may be observed with Fermi LAT in the
near future.}
\label{tab:dwarf}
\end{table}

\begin{figure}[htpb]
\centering
\includegraphics[width=15cm]{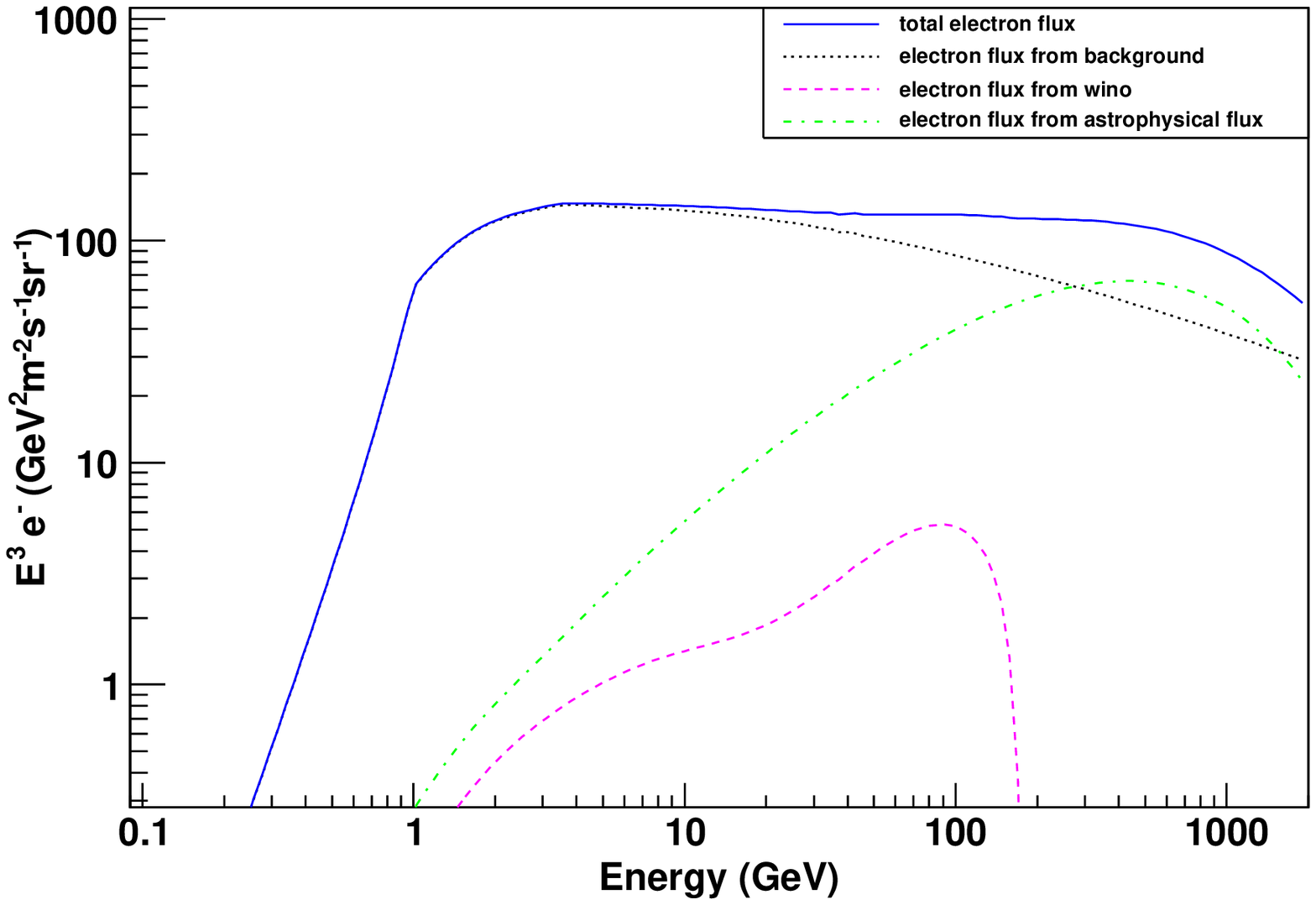} 
\caption{The absolute flux of $e^-$, the solid line is the sum of all the
components, the dot line is the conventional astrophysics background, the
dash line is the wino annihilation signal, and the dash-dot line is the
contribution from the extra flux.}
\label{fig:em_flux}
\end{figure}

\begin{figure}[htpb]
\centering
\includegraphics[width=15cm]{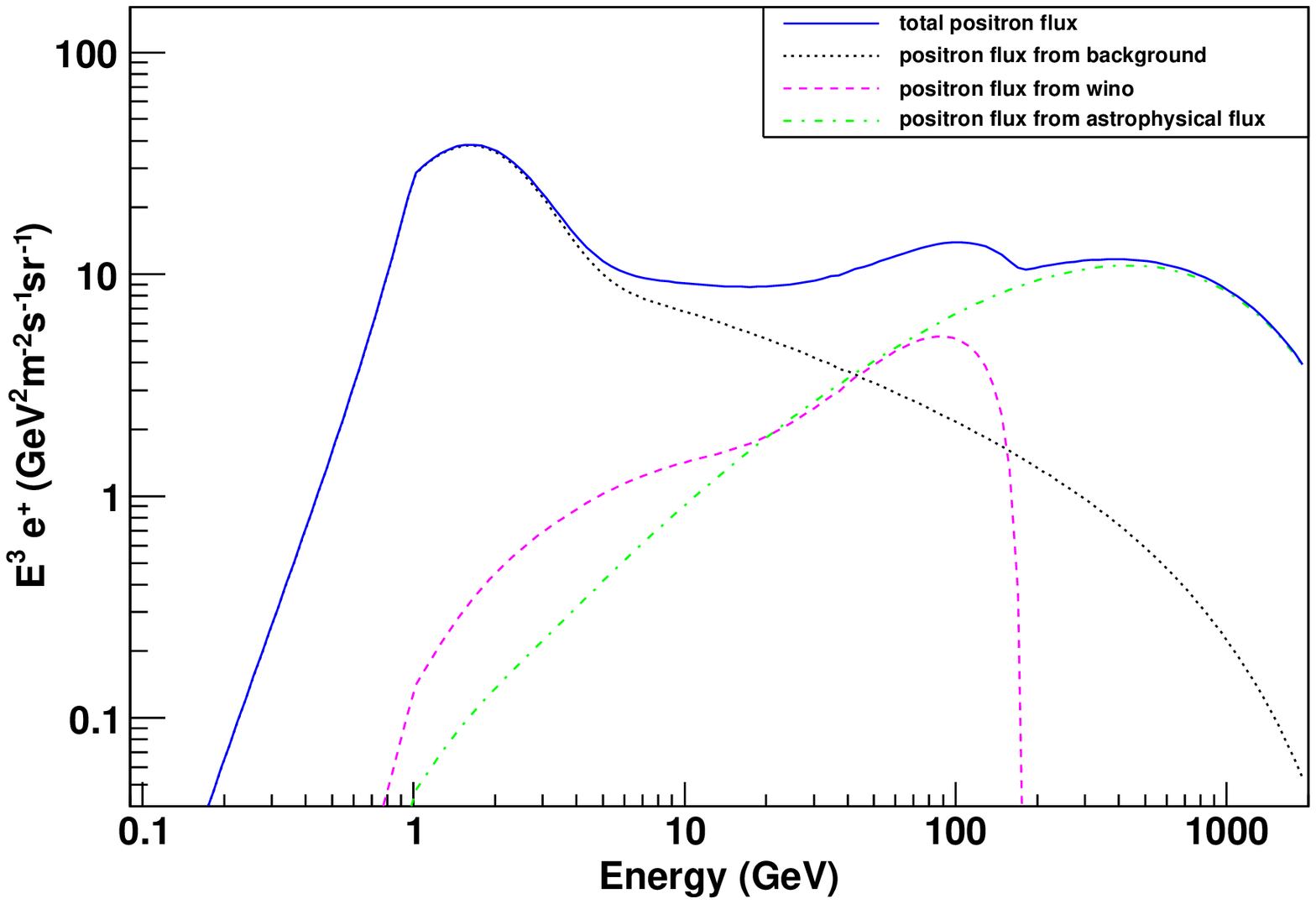} 
\caption{The absolute flux of $e^+$, the solid line is the sum of all the
components, the dot line is the conventional astrophysics background, the
dash line is the wino annihilation signal, and the dash-dot line is the
contribution from the extra flux.}
\label{fig:ep_flux}
\end{figure}

\begin{figure}[htpb]
\centering
\includegraphics[width=15cm]{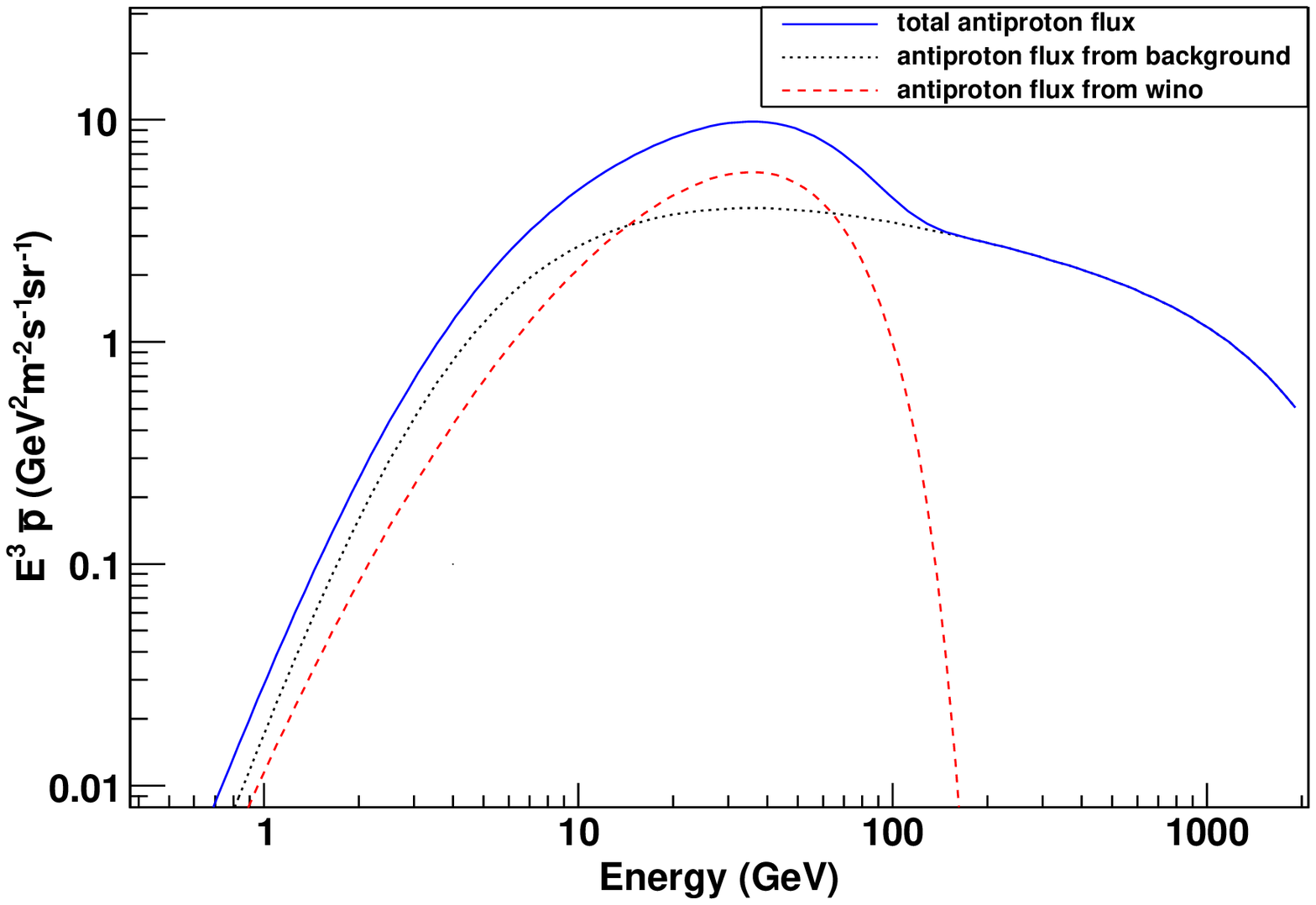} 
\caption{The absolute flux of $\bar{p}$, the solid line is the sum of all the
components, the dot line is the conventional astrophysics background, and the
dash line is the wino annihilation signal.}
\label{fig:pb_flux}
\end{figure}
\clearpage

\section{LHC}
The LHC phenomenology of the wino LSP is fairly well studied because it
originally occurred in the anomaly mediated
context\cite{Randall:1998uk,Moroi:1999zb}, The production rates are
large, $\sim 10 \,\mathrm{pb}$ for charginos and neutralinos. The triggers and
signatures are difficult since the chargino and LSP are approximately
degenerate\cite{Feng:1999fu,Gherghetta:1999sw,Ibe:2006de,Asai:2008sk,Acharya:2008zi},. Probably the main trigger will be the associated gluino
production and decay. Gluino masses in models range from a few to about 10
times the LSP mass, all within the LHC range. We will report on these topics
later.
\section{Summary of tests and comments}
There will be definitive tests of the existence of a wino LSP with the mass
range we consider at the LHC. We
summarize here tests that will occur from astrophysical data and analysis
soon. At this stage we cannot make strong statements about direct detection,
since pure winos have small scattering cross section. The rates are very
sensitive to mixtures of binos and higgsinos, but the present data plus the
propagation uncertainties do not determine the mixtures very well. Also the
contribution from $\chi\chi\rightarrow \gamma\gamma$ and $\chi\chi\rightarrow
Z\gamma$ is not included in this study, since the cross sections, calculated
with DarkSUSY using the analytical formulas given in \cite{Bergstrom:1997fh},
are relatively small ($2.25 \times 10^{-27} \mathrm{cm^3 s^{-1}}$ for
$\gamma\gamma$ and $1.36 \times 10^{-26} \mathrm{cm^3 s^{-1}}$ for $Z
\gamma$ respectively).

Astrophysical tests includes the following. For some the
presence of high energy astrophysical contribution needs to be kept in
mind. When future data is reported we will post updated graphs at
\url{http://wino.physics.lsa.umich.edu}.
\begin{enumerate}
    \item Turnover or flattening of the positron ratio and the positron absolute
        flux with increasing energy. (Figure \ref{fig:ep_ratio},
        \ref{fig:ep_flux}).
    \item The rise in the positron ratio is not due to a decrease in the
        electron flux, which will not decrease faster in the region from 10 -
        200 GeV.
        (Figure \ref{fig:em_flux}).
    \item The $\bar{p}$ rate will turn over with increasing energy. (Figure
        \ref{fig:pb_ratio}, \ref{fig:pb_flux}).
    \item There will be an observable excess in the region below $200
        \,\mathrm{GeV}$ in the diffusive gamma spectrum, by a factor of order $3-4$ from
        wino annihilation (Figure \ref{fig:gamma_fermi_components})
    \item There will be an increase in gammas from the galactic center below
        $200 \,\mathrm{GeV}$ from wino annihilation, almost an order of
        magnitude (Figure \ref{fig:gamma_center})
    \item Effects on synchrotron radiation (WMAP
        haze)\cite{Hooper:2008zg,Cumberbatch:2009ji} and
        recombination\cite{Padmanabhan:2005es,Galli:2009zc,Slatyer:2009yq}
        need further detailed study. Wino annihilation is consistent with the current
        experimental
        constraints\cite{Slatyer:2009yq,Grajek:2008jb,Grajek:2008pg}, though
        barely if all their assumptions are accepted. This may mean the wino
        annihilation is an explanation. Planck data will provide a significant
        test here.
    \item Effects from wino annihilation for dwarf galaxies are probably observable (Table \ref{tab:dwarf})
\end{enumerate}
If wino-like dark matter annihilation is indeed being observed, the
implications are remarkable. We are not only learning what constitutes the dark
matter, it is also the discovery of supersymmetry, and learning that the
universe has a non-thermal cosmological history that can be studied. These
implications in turn favor certain underlying theories.
\section{Acknowledgements}
We are grateful for discussions with Bobby Acharya, Nima Arkani-Hamed, Lars
Bergstr\"{o}m, Elliott Bloom,  Mirko Boezio, Joachim Edsj\"{o},  Rouven Essig,  Phill
Grajek, Emiliano Macchiutti, Patrick Meade, Aldo Morselli, Igor Moskolenko,
Michele Papucci, Cheng Peng, Piergiorgio Picozza, Tomer Volansky, Liantao
Wang, Neal Weiner, and particularly Aaron Pierce who also collaborated on
parts of the research.
S.W. would like to thank KITP, Perimeter Institute, and the University 
of Texas - Austin for hospitality, and also U of T - Austin for 
financial support
under National Science Foundation Grant No. PHY-0455649.
\bibliography{draft}
\end{document}